# RFQ-DTL MATCHING SOLUTIONS FOR DIFFERENT REQUIREMENTS *

D. Raparia, Brookhaven National Laboratory, Upton, NY 11973, USA

## Abstract

The Radio-Frequency Quadrupole (RFQ) has a FODO lattice and Drift-Tube-Linacs (DTL) in general also have FODO lattices. Therefore the natural solution for the matching between these is a FODO lattice. For matching in all three planes one then needs sixteen degrees of freedom. However, different requirements, depending on the applications, may change this solution. For example, in the production environment (like medical and industrial applications), one needs fixed current and fixed beam quality. On the other hand, in the research environment one not only needs all degrees of freedom, but may also want to chop of beam pulse. This paper discusses matching solutions for these different requirements.

## I. Introduction

To provide successful beam delivery from one device to another, one generally needs a beam-line matching section connecting these two devices. The main functions of matching section (MS) are (1) to match the beam into the following device in all phase space, and (2) to provide space for useful (necessary) diagnostics. A good MS should have these functions decoupled. The MS should provide sixteen degree of freedom: ten are machine parameters, namely $\alpha_x, \beta_x, \alpha_y, \beta_y, \alpha_z, \beta_z$ (amplitude function) and $D_x, D_x', D_y, D_y'$ (dispersion function) and six trajectory matching parameters $\Delta x, \Delta x', \Delta y, \Delta y' \Delta z, \Delta z'$. If the RFQ and DTL are in a straight line ( no horizontal or vertical bend), then one only has ten constraints, namely six amplitude functions and four trajectory matching parameters. If the number of variables (*knobs*) is equal to number of constraints, we will call this solution an optimum solution (OS), and if the number of variables is less than the number of constraints, the solution is called over constrained (OCS). Finally if the number of the variables is more than the constraints, the solution will be an under constrained solution (UCS). These solutions may have different lattices, such as FODO [1], [2], [5], [6], [7], [11], FOFODOD [10] or triplet [9].

The choice of solution will depend upon requirements and limitations such as space, emittance growth, funding, etc. The important ingredients which go into the choice of the solution to minimize the emittance growth and particle loss are : (1) Physical beam size; there should not be a sudden change in beam size. In other words, the zero current phase advance per unit length ($\sigma/L$) should have no discontinuity. (2) Space charge forces; the tune depression($\frac{\sigma}{\sigma_0}$ ) should not be too low, (3) Neutralization;

*work performed under the auspices of the U. S. Dept. of Energy.

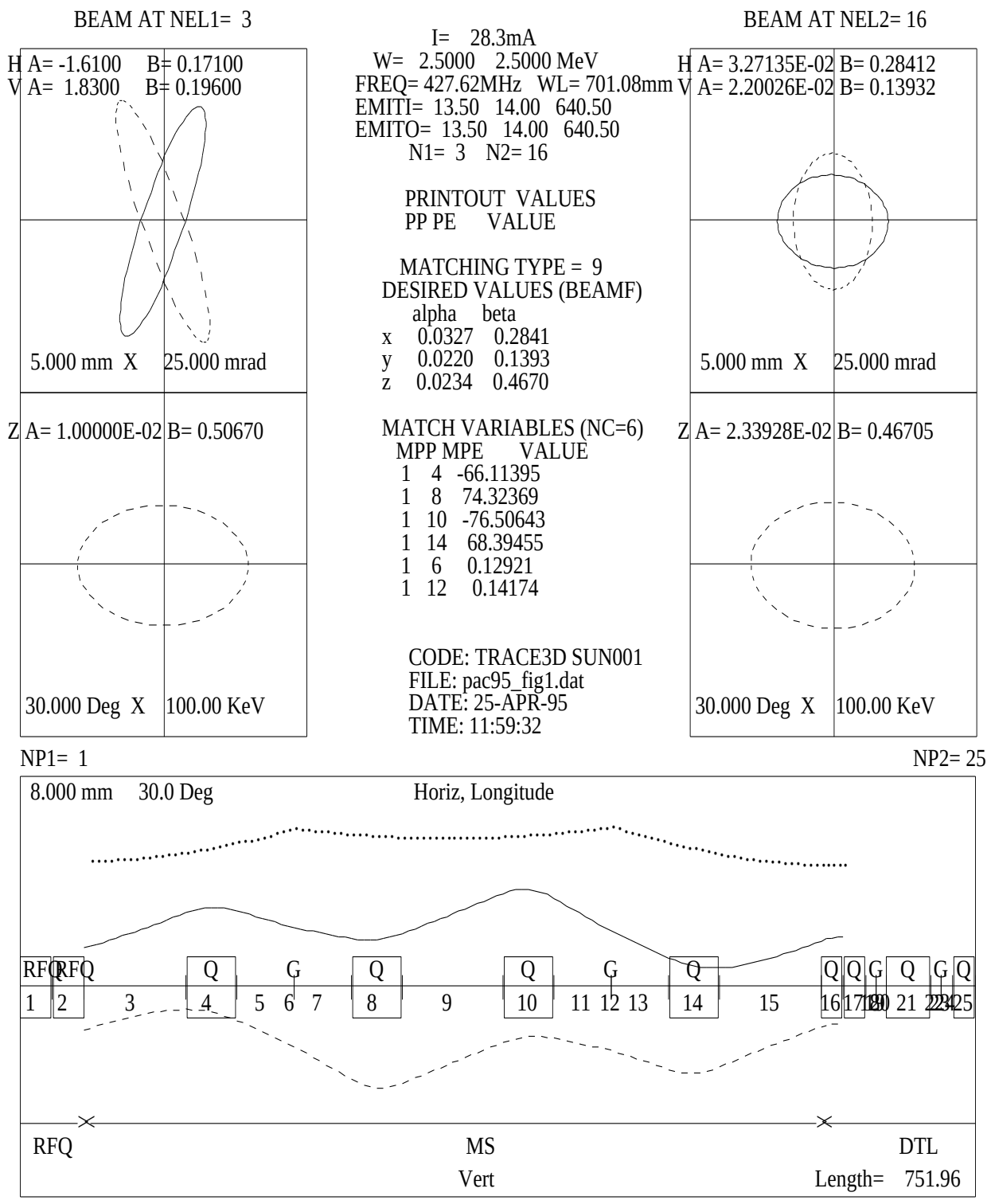

Figure. 1. TRACE3D Beam profiles for OS (SSC MS)

in the case of $H^-$, neutralization should be avoided, particularly when the beam energy is low, (4) Diagnostics; there should be enough space left for necessary diagnostics. The most sensitive errors are the trajectory position matching errors. Emittance growth and particle losses are relatively less sensitive to the amplitude function matching errors. One can estimate the effect of these errors as follows: Position mismatch, $\frac{\sigma^2}{\sigma_0^2} = 1 + \frac{1}{2}\left(\frac{\Delta X_{eq}}{\sigma_0}\right)^2$, and amplitude function mismatch, $\frac{\sigma^2}{\sigma_0^2} = 1 + \frac{1}{2}\left(\frac{\left(\frac{\Delta\beta}{\beta}\right)}{\sqrt{1+\left(\frac{\Delta\beta}{\beta}\right)}}\right)^2$. Where $\Delta X_{eq}^2 = \Delta x^2 + (\alpha\Delta x + \beta\Delta x')^2$, and $\sigma$ and $\sigma_0$ are the rms beam sizes for unmatched and matched beam respectively.

## II. Matching Solutions

RFQs have FODO lattices and, generally, DTLs also have FODO lattices. Therefore, *the natural choice lattice for MS is FODO.* To provide matching in all phase spaces, one needs two FODO cells, two bunchers between quadrupoles, and four steerers [1] (OS). The number of FODO cells may be less than two, depending upon the requirements and limitations (OCS) [8] [6]. The number of cells may also be more than two [2], in order to provide extra constraints such as space for chopper, bending magnet

| Quadrupole No. | 5 cm | 7 cm | 9 cm |
|---|---|---|---|
| Q1 (T/m) | -125.01 | -126.47 | -126.80 |
| Q2 (T/m) | 149.63 | 155.41 | 141.82 |
| Q3 (T/m) | -125.82 | -108.06 | -38.22 |
| Q4 (T/m) | 140.29 | 112.07 | 7.581 |

Table I
DTL First Four Quad. Grad. for 5, 7, 9 cm, MS(OCS)

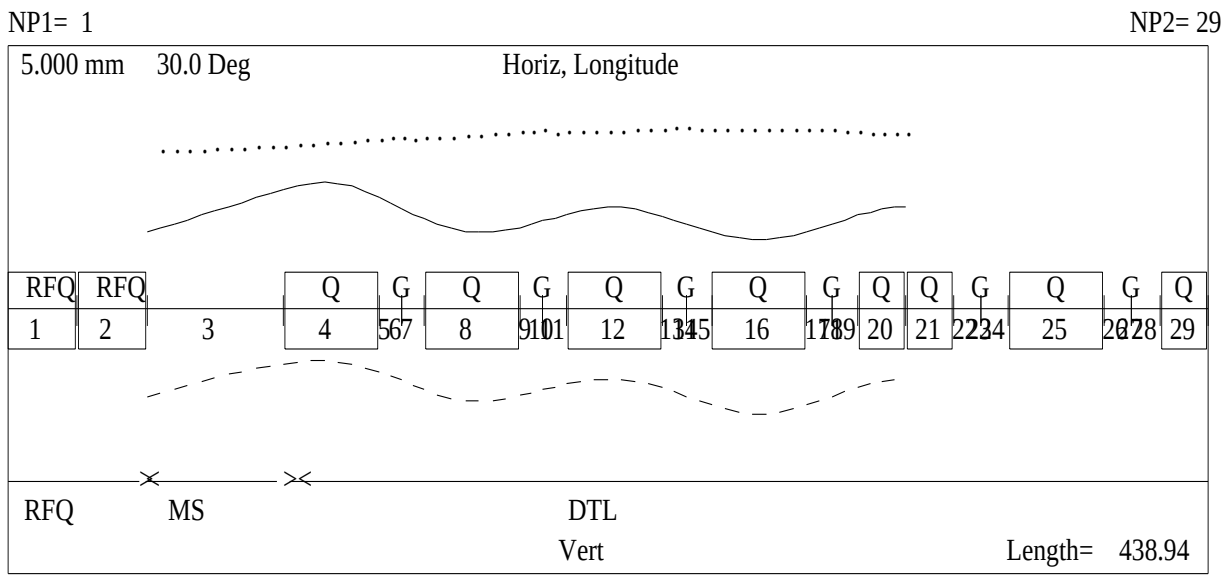

Figure. 2. TRACE3D Beam profiles for 5 cm MS(OCS)

[5], etc (UCS).

In this section we will consider solutions of each type. We will use the SSC RFQ [4] and SSC DTL Tank1 [3]. The output energies of the RFQ and Tank1 are 2.5 MeV and 13.4 MeV, respectively and the noninal current is 25 mA.

## *A. Optimum Solution (OS)*

Figure 1 shows the TRACE3D beam profiles for the SSC MS. This MS has four variable PMQs and two bunchers, to provide six variables for amplitude function. The variable PMQs were also movable transversely to provide four steering variables to match trajectory. It had enough space to provide diagnostics. The phase advance per $\beta\lambda$ at the end of RFQ is 22.40 deg, and at the beginning of DTL was 20.85 deg. The tune depression in this section is 0.92.

## *B. Over Constrained Solution (OCS)*

In the production environment, one needs reliability rather than flexibility. The fewer the variables, the better the reliability. Partial or full matching may be accomplished by altering a few end cells of the RFQ and the first few cells of the DTL.

In the following examples we have only drift lengths of 5,7,and 9 cm between the RFQ and the DTL and have used the first four quadrupoles in the DTL for the matching in the following cells, The first four quadrupole gradients for these cases are given in Table I. TRACE3D profiles for case of 5 cm MS(OCS) are shown in Figure 2.

## *C. Under Constrained Solution (UCS)*

In this situation, one might have to accommodate other constraints. For example, one might have to chop the beam pulse length [9], or the DTL lattice is not FODO [10], or $\sigma_0/L$ is quite different in the RFQ and DTL [7].

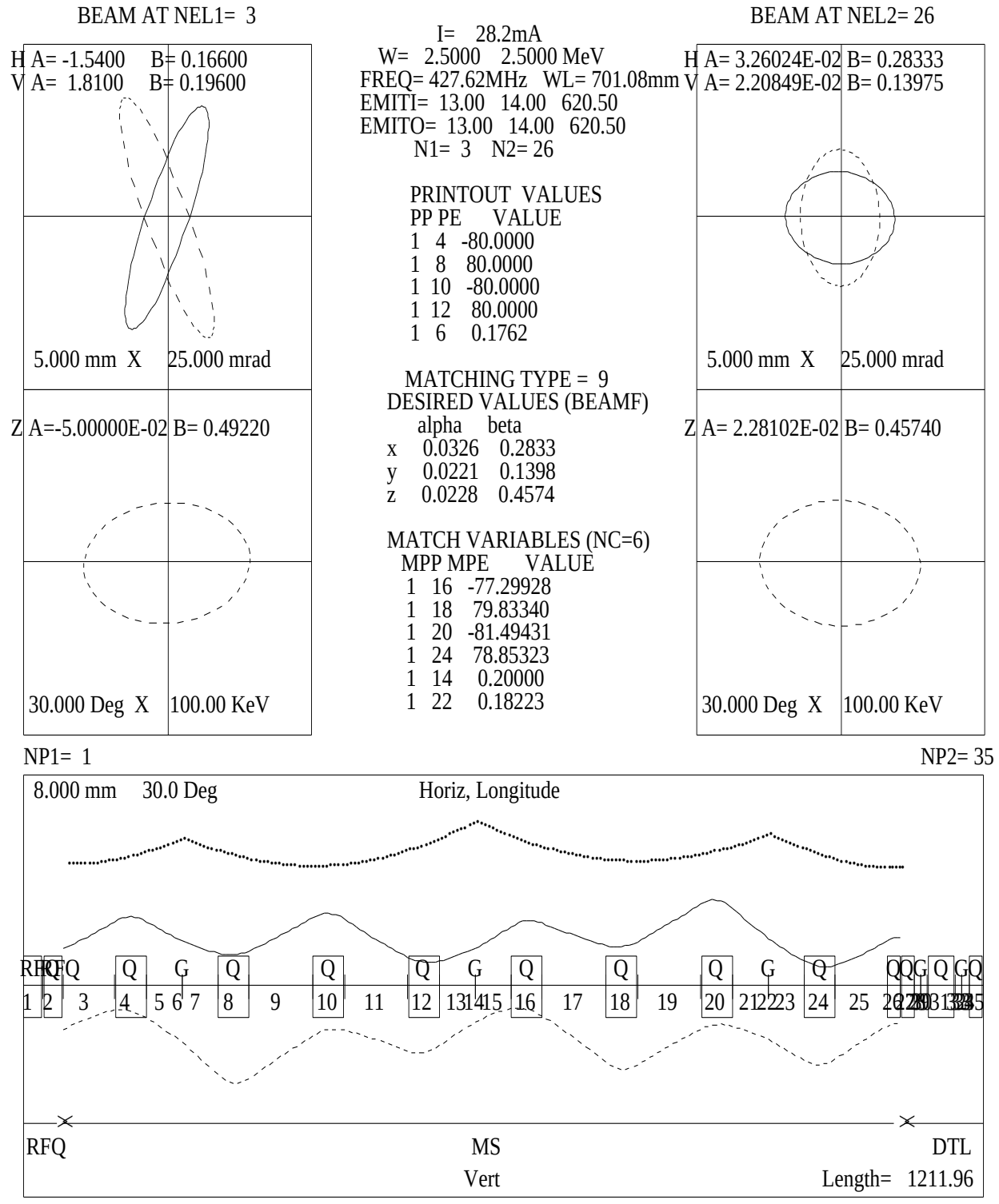

Figure. 3. TRACE3D Beam Profiles for UCS (FODO).

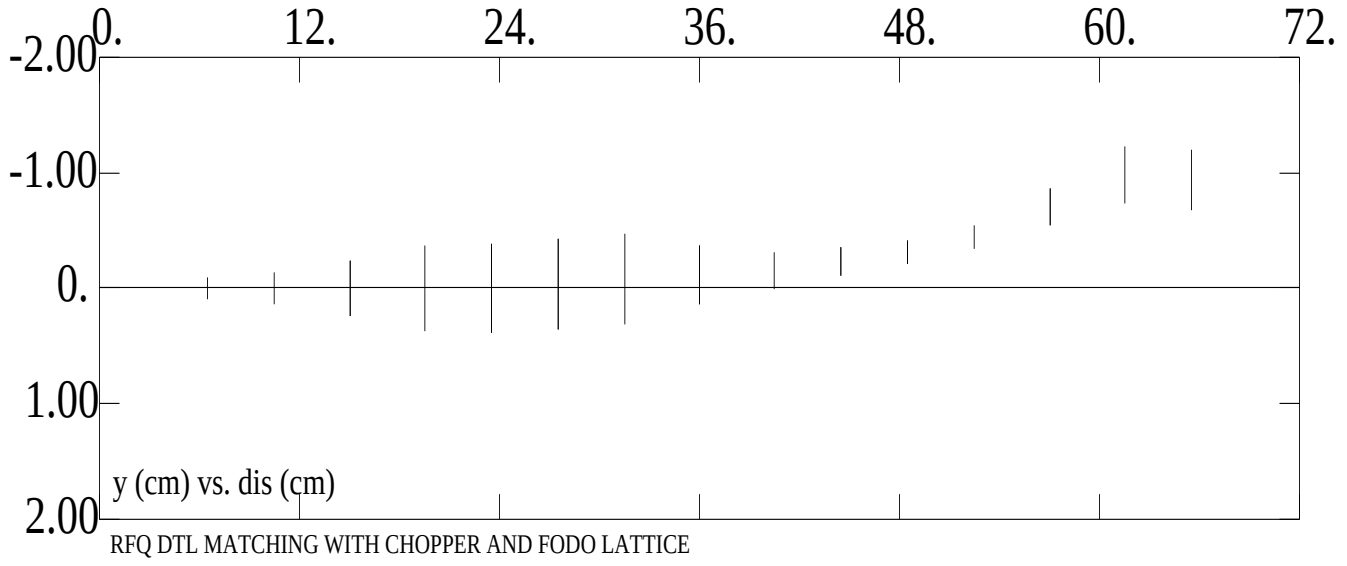

Figure. 4. PARTRACE Beam profile for UCS (FODO) with chopper.

We consider an RF chopper to chop the beam. We have tried two solutions namely, (1) FODO lattice where choppers are placed between quadrupoles and, (2) triplets to provide long drift space for one chopper. In the first case, a systematic search was made for minimum number of choppers which could kick the beam centroid at least 0.8 cm (beam pipe radius). The TRACE3D beam profiles are shown in Figure 3 for the FODO solution which was optimize for the minimum emittance growth. In this case we have four choppers, having plates which are 6 cm long and 2.54 cm apart. The chopper pulse is about 5kV at few MHz. These choppers are located at element numbers 9, 11, 17 and 19 in Figure 3. This arrangment could kick the beam centroid 1 cm off axis as shown in Figure 4.

For the second solution, to provide 30 cm long drift space with beam size less than 0.8 cm, we have used two triplets, and four quadrupoles to bring the beam size slowly to match to the DTL. Again this solution is also optimized for the minimum emittance growth. Figure 5 shows the TRACE3D profiles for the triplet lattice. Again, the chopper operating parameters are same as above, but, instead four 6 cm long plates, it uses one 30 cm long plate, corresponding to element number 9 in Figure 5. Figure 6 shows the beam profiles through this section. In this case also the beam centroid is kicked 1 cm off axis.

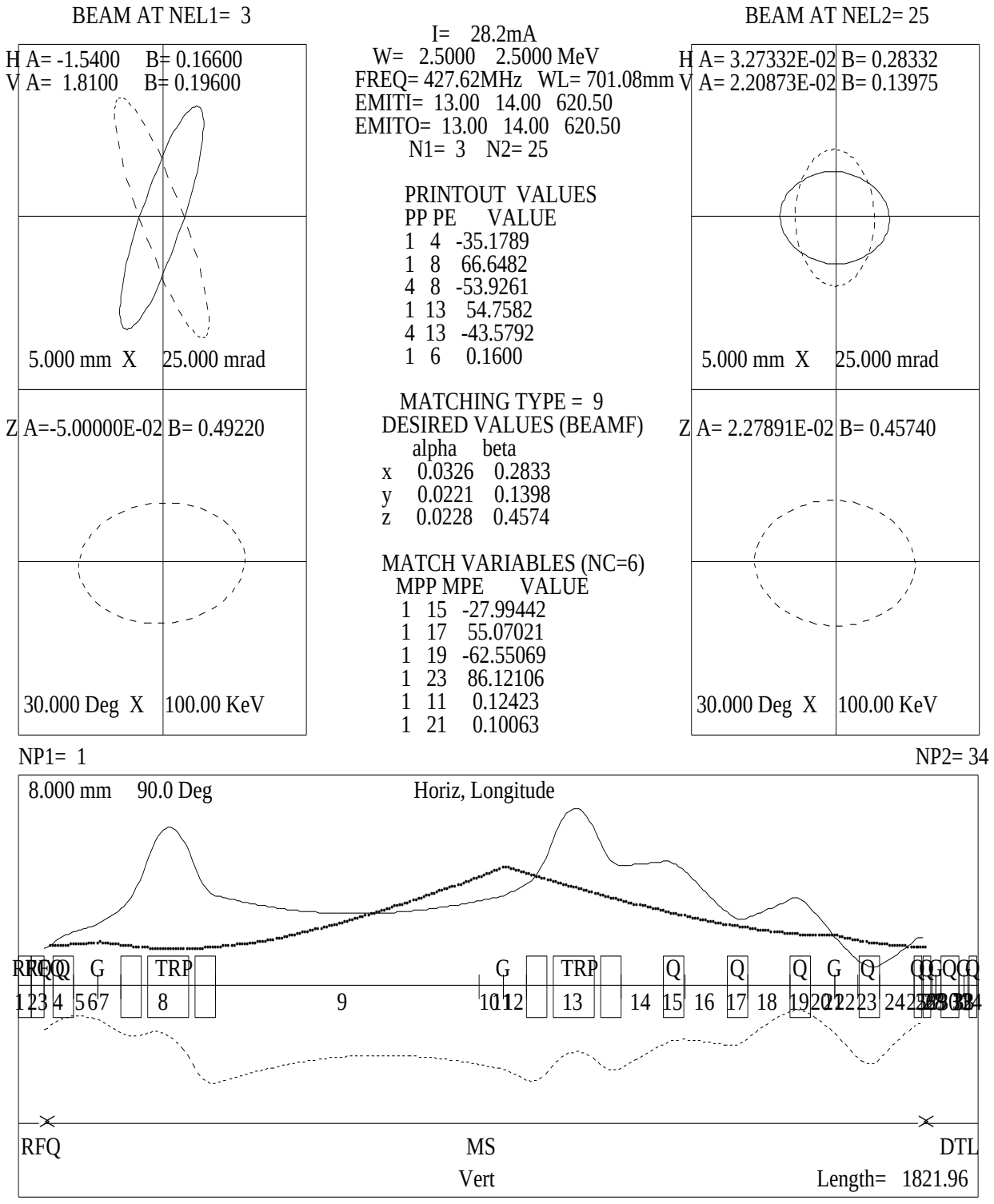

Figure. 5. TRACE3D Beam Profiles for UCS (Triplet).

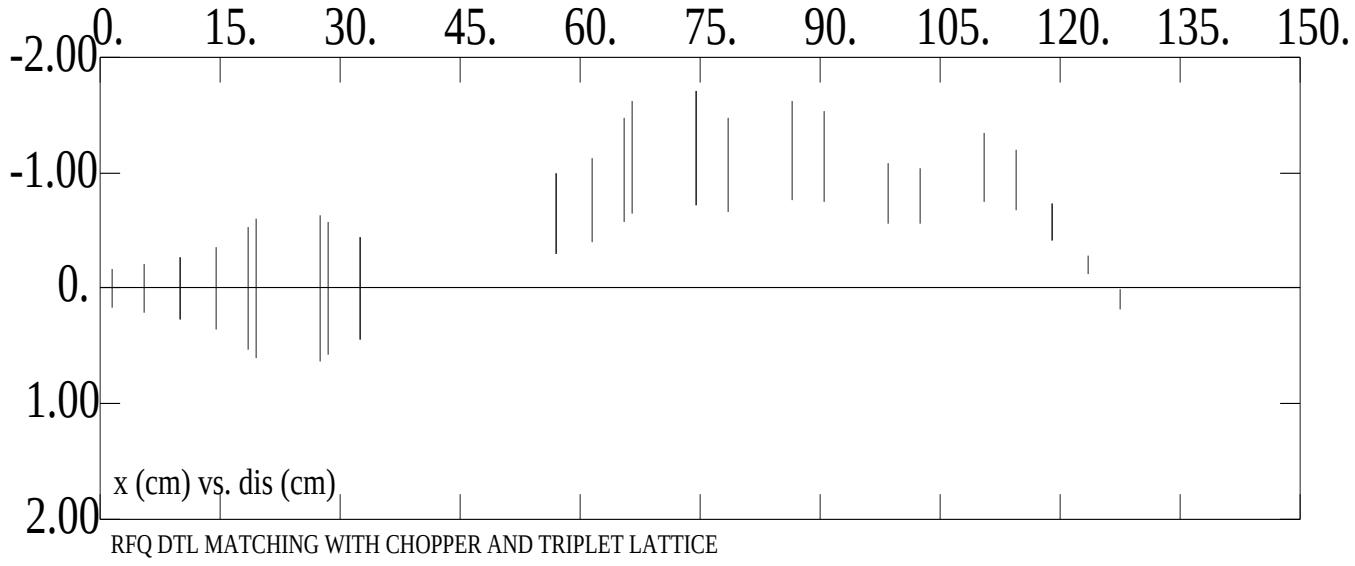

Figure. 6. PARTRACE Beam Profiles for UCS (Triplet) with chopper.

## III. Conclusions

We have also done PARMTEQ and PARMILA calculations for these cases. 1000 macro-particles were used to form a matched beam into the RFQ. The same particles were followed in the MS and DTL. Table II shows the output emittances at the end of TANK 1. In MS and DTL no particle loss occurred except in the case of UCS triplet where particle loss is about 0.5%. The emittance growth in casees of OS, OCS (5 cm) and UCS (FODO) are reasonable.

| Solutions | $\epsilon_x$ | $\epsilon_y$ | $\epsilon_z$ | $\Delta\epsilon_t$ | $\Delta\epsilon_z$ |
|---|---|---|---|---|---|
| RFQ | 0.189 | 0.208 | 0.124 | - | - |
| OS | 0.202 | 0.219 | 0.124 | 5.% | 0.% |
| OCS (5 cm) | 0.194 | 0.214 | 0.149 | 4.% | 20.% |
| OCS (7 cm) | 0.213 | 0.215 | 0.161 | 10.% | 30.% |
| OCS (9 cm) | 0.347 | 0.215 | 0.188 | 44.% | 52.% |
| UCS (FODO) | 0.208 | 0.217 | 0.139 | 6.% | 12.% |
| UCS (Triplet) | 0.193 | 0.243 | 0.150 | 12.% | 21.% |

Table II

DTL Tank 1 output normalized rms emittances. $\epsilon_x$,$\epsilon_y$ are in units of $\pi$ mm-mrad, $\epsilon_z$ is in units of $\pi$ MeV deg. $\Delta\epsilon_t$ is the average emittance growth in x and y plane with respect to the RFQ output.

## IV. Acknowledgement

We would like to thank J. Alessi and A. Kponou for their valuable suggestions and discussions.

## References

[1] R. C. Sethi, *et al*, " Design of the RFQ-DTL Matching Section for the SSCL Linac", p 482, Proceeding of the 1992 Linear Accelerator Conference, AECL-10728, August 1992, Ottawa, Ontario, Canada.

[2] O. R. Sander, *et al* , " Commissioning the GTA Accelerator", p 535, *ibid.*

[3] D. Raparia, *et al*," SSC Drift-Tube Linac Design ", p 199, *ibid.*

[4] T. S. Bhatia, *et al*," Beam Dynamics Design of an RFQ for the SSCL", p 1884, Proceeding of the IEEE Particle Accelerator Conference, May 1991, San Francisco, CA.

[5] D. Raparia, " RFQ-DTL Matching Section", Preliminary Design Review for SSC Linac", January 29, 1991.

[6] E. Boltezar, *et al*, " Experimental RFQ As Injector To The CERN Linac 1", p 302, Proceeding of the 1981 Linear Accelerator Conference, AECL-10728, October 1981, Santa Fe, New Maxico, USA.

[7] M. Weiss, " The RFQ2 Complex: the Future Injector to CERN Linac2", p 539, Proceeding of the 3rd European Particle Accelerator Conference, March 1992, Berlin.

[8] R. Hamm, Private Comunication.

[9] J. Alessi, *et al*, " The AGS $H^-$ RFQ Preinjector", p 196, Proceeding of the 1988 Linear Accelerator Conference, CEBAF-Report-89-001, October 1988, Newport News, Virginia.

[10] R. W. Garnett and P. Smith, "Design of a Current-Independent Matching Section for APDF", p 107, Proceeding of the 1994 Linear Accelerator Conference, August 1994, Tsukuba, Japan.

[11] Takao Kato, " Design of Beam-Transport Line Between the RFQ and the DTL for the PHP 1-GeV Proton Linac", p 59, *ibid.*